\documentclass[12pt,epsf]{article}
\usepackage{amsmath}
\usepackage{amssymb}
\usepackage{theorem}
\usepackage{latexsym}
\usepackage{psfig}
\usepackage{graphicx,color}

\newcommand{\rem}[1]{}

\catcode`\@=11

\title{\bf
On propagation failure in 1 and 2 dimensional excitable media}
\author 
{
{\em Georg A. Gottwald$^1$ and Lorenz Kramer$^2$}\\
{\em $^1$ School of Mathematics \& Statistics, University of
Sydney,}\\
{\em NSW 2006, Australia.}\\
{\small gottwald@maths.usyd.edu.au}\\
{\em $^2$ Physikalisches Institut, Universit\"at Bayreuth, }\\
{\em Universit\H atstra{{\ss}}e 30, D-95440, Bayreuth, Germany.}\\
{\small lorenz.kramer@uni-bayreuth.de}\\
}
\date{}

\begin{document}

\begin{titlepage}
\setcounter{page}{1}

\vfill
\maketitle

\begin{abstract}
\noindent  
We present a non-perturbative technique to study pulse dynamics in
excitable media. The method is used to study propagation failure in
one-dimensional and two-dimensional excitable media. In
one-dimensional media we describe the behaviour of pulses and
wave trains near the saddle node bifurcation, where propagation
fails. The generalization of our method to two dimensions captures the
point where a broken front (or finger) starts to retract. We obtain
approximate expressions for the pulse shape, pulse velocity and
scaling behavior. The results are compared with numerical simulations
and show good agreement.
\end{abstract}
\vfill

\end{titlepage}


{\bf{ Excitable media are often found in biological and chemical
systems. Examples of excitable media include electrical waves in
cardiac and nerval tissue \cite{WinfreeBook,Davidenko}, cAMP waves in
slime mold aggregation \cite{dictyostelium} and intracellular calcium
waves \cite{calcium}. Excitable media support localized pulses and
periodic wave trains. In 2 dimensions rotating vortices (or spirals)
are possible \cite{Winfree}. The critical behavior of pulses, wave
trains and spirals, i.e.propagation failure, is often associated with
clinical situations. The study of spiral waves is particularly
important as they are believed to be responsible for pathological
cardiac arrhythmias \cite{Chaos}. Spiral waves may be created in the heart through
inhomogeneities of the properties of the cardiac tissue.\\ We
investigate critical behavior relating to these 3 wave types. We
develop a non-perturbative test function method which allows to study
the bifurcation behavior of critical waves. In particular, we study
under what conditions a broken front will sprout and develop into a
spiral wave or retract. Analytical formulas for the growing velocity
of a broken front are given. For wave trains we provide a time
dependent extension which supports a Hopf bifurcation which is also
observed in numerical simulations of excitable media. This seems to be
related to alternans \cite{Nolasco,Karma_A}, which also are discussed
in the context of cardiac electric pulse propagation. The methods and
results are general, and can be applied to other excitable media.  }}


\section{Introduction}
Many chemical and biological systems exhibit excitability.  In small
(zero-dimensional) geometry they show threshold behavior, i.e. small
perturbations immediately decay, whereas sufficiently large
perturbations decay only after a large excursion. One dimensional
(1D) excitable media support travelling pulses, or rather, periodic
wave trains ranging in wavelength $L$ from the localized limit $L\to
\infty$ to a minimal value $L_c$. Pulses and wavetrains are best-known from nerve
propagation along axons. In 2D one typically observes spiral
waves. Spirals have been observed for example in the auto-catalytic
Belousov-Zhabotinsky reaction \cite{Winfree}, in the aggregation of
the slime mold dictyostelium discoideum \cite{dictyostelium} and in
cardiac tissue \cite{Davidenko}.

For certain system parameters the propagation of pulses, wave trains
or the development of spiral waves may fail (see for example
\cite{synergetics,Jahnke}). The analytical tools employed to describe
these phenomena range from kinematic theory \cite{Zykov,Mikhailov1},
asymptotic perturbation theory \cite{Meron,Karma1,Karma2} to dynamical
systems approaches \cite{barkley94,Melbourne}. However, no theory
exists which describes propagation failure using only equation
parameters, and which reproduces the behavior close to the bifurcation
point. For example, asymptotic perturbation theory fails to describe
the square-root scaling behavior of the amplitude and the pulse
velocity with respect to the bifurcation parameter at the bifurcation
point. In kinematic theory results are not given entirely in terms of
the system parameters. In this paper we develop a non-perturbative
method to study propagation failure and compare the results with
numerical simulations.\\

Most theoretical investigations are based on coupled
reaction-diffusion models. We follow this tradition and investigate a two-component,
two-dimensional excitable medium with an activator $u$ and a
non-diffusive inhibitor $v$ described by
\begin{eqnarray}
\label{barkley}
\partial_t u &=& D \Delta u +{\cal{F}}(u,v),\quad {\cal{F}}(u,v) = u(1-u)(u-u_s-v) \nonumber \\
\partial_t v &=&  \epsilon \ (u- a\  v)\  .
\end{eqnarray}
This is a reparametrized version of a model introduced by Barkley
\cite{barkley91}. We expect our method to be independent of the
particular model used. Note that the diffusion constant $D$ is not a
relevant parameter as it can be scaled out by rescaling the length.

This model incorporates the ingredients of an excitable system in a
compact and lucid way. Thus, for $u_s>0$ the rest state $u_0=v_0=0$ is
linearly stable with decay rates $\sigma_1=u_s$ along the activator
direction and $\sigma_2=\epsilon a$ along the inhibitor
direction. Perturbing $u$ above the threshold $u_s$ (in 0D) will lead
to growth of $u$.  In the absence of $v$ the activator would saturate
at $u=1$ leading to a bistable system.  A positive inhibitor growth
factor $\epsilon$ and $a>0$ forces the activator to decay back to
$u=0$. Finally also the inhibitor with the refractory time constant
$(\epsilon\ a)^{-1}$ will decay back to $v=0$. For $a>1/(1-u_s)$ the
system is in zero-dimensional systems no longer excitable but instead
bistable.
This relaxation behaviour in the 0D system for super-threshold
perturbations gives rise to pulse solutions in the 1D (and 2D)
case. The relaxation mechanism mediated by the inhibitor forces the
pulse solution to decay in its back. Hence we observe pulses in
excitable media and not fronts.

The Barkley model is a variant of the class of 2-component
Fitzhugh-Nagumo models. In the traditional Fitzhugh-Nagumo models
${\cal{F}}(u,v)$ in Eq.(\ref{barkley}) is replaced by
${\cal{F}}_{FN}(u,v)= u(1-u)(u-u_s)-v$. Thus the nullclines
${\cal{F}}(u,v)=0$, which in the traditional Fitzhugh-Nagumo models
are cubic polynomials, are replaced by straight lines in the Barkley
model. Most of the qualitative behavior in the relevant parameter
ranges is unchanged by this. Whereas the Barkley model is
computationally more efficient and also analytically better tractable,
the traditional Fitzhugh-Nagumo models display a feature which makes
them more realistic for the discription of excitable media in biology:
the activator experiences an undershoot below its equilibrium value
and slow decay in the tail region of a pulse. We expect that for the
phenomena discussed in this paper the difference only leads to
quantitative changes (indeed, we have done some tests to verify this
assertion).

In order to study pulse propagation in 1D it is useful to first
consider the case of constant $v$. The resulting bistable model is
exactly solvable \cite{Keener} and the pulse velocity is
$c_f(v)=\sqrt{\frac{D}{2}}[1-2(u_s+v)]$. Hence, excitability requires
that $u_s$ is below the stall value $\frac{1}{2}$. The quantity
$\Delta=\frac{1}{2}-u_s$ characterizes the strength of excitability
and $c_f(0)$ coincides with the solitary pulse velocity for $\epsilon
\to 0$.

Clearly, for $u_s<u_c=\frac{1}{2}$ and not too large $a$, pulse
propagation fails for $\epsilon$ larger than some $\epsilon_c$. The
critical growth factor $\epsilon_c$ describes the onset of a
saddle-node bifurcation
\cite{Zykov,Meron}. The saddle node can be intuitively
understood when we consider the activator pulse as a heat source, not
unlike a fire-front in a bushfire. Due to the inhibitor the width of
the pulse decreases with increasing $\epsilon$. Hence, the heat
contained within the pulse decreases. At a critical width, or a
critical $\epsilon$, the heat contained within the pulse is too small
to ignite/excite the medium in front of the pulse.

For periodic wavetrains the saddle node depends on the wavelength.
The pulses run into the inhibitor field of their respective preceding
pulse. Hence, propagation failure for periodic wave trains is
controlled by the decay of the inhibitor and propagation is only
possible when the inter-pulse distance $L$ becomes larger than a
critical wavelength $L_c$. Note that $L_c$ diverges for $a \to 0$ when
the decay rate of the inhibitor $\sigma_2$ vanishes.\\

In previous analytic works on one-dimensional pulses the limit of
small $\epsilon$ was considered \cite{Zykov,Meron}. Then the solution
of the activator $u$ is well separated in two flat plateau regions
with $u\approx 1$ and $u=0$ which are separated by a steep narrow
front. This approach does not capture the square-root behavior of
$c_0(\epsilon)$ close to the saddle-node bifurcation point. Close to
the bifurcation point the solution resembles rather a bell-shaped
pulse than a plateau. In this paper we will be concerned with the
behavior near the bifurcation at $\epsilon_c$ and make explicit use of
the observed bell-shape form of the solution. This allows us to
describe the scaling behavior close to the bifurcation in terms of the
equation parameters.

In 2 dimensions spiral waves are observed. Spirals can be created in
excitable media from a finger, i.e. a 1D pulse which is extended in
the second dimension and has one free end. Fingers may be created due
to inhomogeneities in the excitability of the system
\cite{Krinsky}. At long times, the free end will either sprout or
retract depending on the growing velocity $c_g$ being positive or
negative. If $c_g>0$, the tip of the finger will sprout into the fresh
medium, and in particular it sprouts and curves backwards. This causes
a non-vanishing curvature at the tip of the finger. Due to the
increased curvature the finger tip is slower than the flat part of the
front further away from the tip. Thus, the extending part far from the
tip will curl up. This leads to the formation of a spiral with the
free end at its core. The criterion $c_g>0$ is therefore a necessary
criterion for spiral formation.

The transition to retraction always occurs before the 1D propagation
failure. It is harder to tackle analytically. General and universal
dynamical system approaches exploiting only the Euclidean symmetry of
excitable media describe the transition as a drift bifurcation. These
model independent theories give an explanation for the divergence of
the core radius at the bifurcation point and also explain why a finger
at the bifurcation point is translating with finite speed, i.e. that
the transition to retracting fingers always occurs before the 1D
propagation failure
\cite{Melbourne}. Unfortunately, it has the unphysical result that 
at the bifurcation point $\epsilon_g$ a spiral changes its sense of
rotation \cite{Melbourne}. Whereas these theories treat the spiral as
a global solution of the underlying equations and see the transition
to spiral waves as a pitchfork or drift bifurcation
\cite{barkley94,Melbourne}, we take a local approach and describe the
transition not as a bifurcation but as a quantitative change in the
velocity of the finger, analogous to a Maxwell point in a first-oder
phase transition. Asymptotic techniques in the limit $\epsilon \ll 1$
have also been performed for this problem
\cite{Karma1,Karma2} and produced an analytical expression for the
onset of retraction for small $\epsilon$. Moreover, the authors were
able to go one step further, in a detailed and sophisticated
asymptotic analysis, and described the onset of meandering. In this
paper we go beyond the restriction of small $\epsilon$ and propose a
more intuitive approach to the problem of the growing velocity which
nevertheless gives good agreement with the numerics.\\

In the following we assume given excitability parameters $a$, $u_s$
and take $\epsilon$ as the bifurcation parameter. We will be looking
for solutions that move with velocity $c_0$ in the $x$ direction and
may grow (or retract) with velocity $c_g$ in the $y$ direction. Thus
we rewrite Eqs. (\ref{barkley}) in a frame moving with velocity
$(c_0,\ c_g)$ as
\begin{eqnarray}
\label{barkleyT1}
D (\partial_{x}^2+\partial_{y}^2) u + c_0 \partial_{x} u +
c_g \partial_{y} u+ {\cal{F}}(u,v)&=&0  \\
\label{barkleyT2}
c_0 \partial_{x} v  +c_g \partial_{y} v+ \epsilon(u- a\ v)&=&0 \; \; .
\end{eqnarray}
\vskip 5pt


\noindent
\section {Pulse propagation in one-dimensional excitable media}
\vskip 5pt 
We first look for pulse and wavetrain solutions that do not depend on
$y$. The reason for the failure to describe the pulse properties at
$\epsilon_c$ within the framework of asymptotics employing the
smallness of $\epsilon$ is due to the fact that at $\epsilon_c$ the
pulse shape for $u$ cannot be separated into a steep narrow front and
a flat plateau. Hence, asymptotic techniques such as inner and outer
expansions are bound to fail. Instead the pulse has the shape of a
rather symmetric bell-shaped function (see Fig. 1). In the following
we make explicit use of the shape of the pulse close to the critical
point and parametrize the pulse appropriately; a method reminiscent of
the method of collective coordinates in the studies of solitary waves
\cite{Scott}.

We choose $u$ of the general form
\begin{eqnarray}
\label{ansatz}
u(x)=f_0 U(\eta) \qquad {\rm{with}} \qquad \eta= w x \; ,
\end{eqnarray}
where $U(\eta)$ is chosen as a symmetric, bell-shaped function, for
example a Gaussian, of unit width and height. Hence, we restrict the
solutions to a subspace of bell-shaped functions $U(\eta)$ which is
parametrized by the amplitude $f_0$ and the inverse pulse width
$w$. The aim of our method is to determine the so far undetermined
parameters. This is done by minimizing the error made by the
restriction to the subspace defined by (\ref{ansatz}).

We avoid further uncontrolled approximations and solve for the
inhibitor field $v$ explicitly
\begin{eqnarray}     \label{vper}
v(\eta)&=& f_0 \Theta V(\eta)\  \qquad {\rm{with}} \qquad 
V(\eta)=e^{a \Theta \eta}\left(V^\star  - \int_{\frac{L}{2}w}^{\eta}d\eta^\prime
e^{-a \Theta \eta^\prime} U(\eta^\prime) \right) \; ,
\end{eqnarray}
where
\begin{eqnarray}     \label{theta}
\Theta=\epsilon/(c_0 w)\; .
\end{eqnarray}
$V^\star$ is determined via the periodic boundary condition
$v(-wL/2)=v(wL/2)$ and is
\begin{eqnarray}   \label{Vstar}
V^\star&=&   \frac{1}{2\sinh (a \Theta \frac{L}{2}w)}
 \int_{-\frac{L}{2}w}^{ \frac{L}{2}w }d\eta^\prime
e^{a \Theta (\eta^\prime-\frac{L}{2}w)} U(\eta^\prime) \; .
\end{eqnarray}
We assume that the width $w^{-1}$ is small compared to the distance
between two consecutive pulses $L$. in the temporal domain this means
that the time scale for the decay of the inhibitor is much longer than
the activator pulse width. This assures that the activator field of
consecutive pulses is well separated and only the inhibitor field
overlaps, and that the interaction between pulses is mediated only
through the inhibitor. Otherwise we would have to choose periodic
functions $U(\eta)$. However, in this case we can now replace the
limits of integration $\pm \frac{L}{2} w$ by $\pm \infty$.  For the
isolated pulse, i.e. when $L \to \infty$, $V^\star$ vanishes.

We now determine the parameters $f_0$ and $w$ by projecting 
Eq. (\ref{barkleyT1}) onto the tangent space of the restricted
subspace defined by (\ref{ansatz}). The tangent space is spanned by
$\partial u / \partial {f_0}=U$ and $\partial u / \partial w=\eta
U_\eta$. This assures that the error made by restricting the solution
space to the test functions is minimized. To achieve this, we multiply
Eq.  (\ref{barkleyT1}) with the basis functions of the tangent-space
$U$ and $\eta U_\eta$, integrate over the $\eta$-domain and require
the projection to vanish, i.e.
\begin{eqnarray}
\label{projectionsU}
\langle Dw^2u_{\eta \eta}+u(1-u)(u-u_s-v) |U \rangle_{u=f_0U(\eta)}=0\\
\label{projectionsxUx}
\langle Dw^2u_{\eta \eta}+u(1-u)(u-u_s-v) |\eta U^\prime
\rangle_{u=f_0U(\eta)}=0\; ,
\end{eqnarray}
where the brackets indicate integration over the whole $\eta$-domain.
The terms proportional to the velocity $c_0$ vanish.

The resulting equations can be combined to give, at fixed $\Theta$ and
$a$, a quadratic equation for $f_0$ with two solutions $f_{0\pm}$
which describe the stable and unstable branch, respectively (see below
for a subtle issue at the saddle node). We obtain
\begin{eqnarray}     \label{u0}
A f_0^2+B f_0+C=0\; ,
\end{eqnarray}
where
\begin{eqnarray}
&A=\frac{3}{4}\langle U^4 \rangle 
     - \frac{5\Theta}{6}\langle U^3 V \rangle
     - \frac{a \Theta^2}{3}\langle \eta U^3 V \rangle \; , \nonumber\\
&B=-\frac{5}{6}(1+u_s)\langle U^3 \rangle 
       + \Theta\langle U^2 V \rangle
       + \frac{a \Theta^2}{2}\langle \eta U^2 V \rangle\ ,  \quad
C=u_s\langle U^2 \rangle \; . \label{coeffs}
\end{eqnarray}
The corresponding inverse width parameters $w_{\pm}$ for the stable
and unstable branch are given by
\begin{eqnarray}
\label{w}
w^2=\frac{1}{D\langle U_{\eta}^2 \rangle}
    [
f_0^2(-\langle U^4 \rangle 
     + \Theta\langle U^3 V \rangle)
+ f_0 ((1+u_s)\langle U^3 \rangle 
      - \Theta\langle U^2 V \rangle)
-u_s\langle U^2 \rangle
]
\; .
\end{eqnarray}
The velocity $c_0$ can now be determined in the standard way by
multiplying Eq. (\ref{barkleyT1}) by $u_x$ and integrating over $x$.
We obtain
\begin{eqnarray}     \label{c0}
c_0&=&
\frac{\int_{-\infty}^\infty{\cal{F}}(u,v)u_x\;
dx}{\int_{-\infty}^\infty u_x^2 dx} \nonumber \\ 
&=&-\frac{f_0 \Theta}{w \langle U_{\eta}^2\rangle}
[\frac{f_0}{3}(\langle U^4\rangle
- a\Theta \langle U^3V\rangle)
- \frac{1}{2}(\langle U^3\rangle - a\Theta \langle U^2 V\rangle)]\; .
\end{eqnarray}
Finally, we can determine $\epsilon$ from $\epsilon=c_0 w \Theta$.
Multiplying (\ref{c0}) by $w$ one sees that $\epsilon$ can be computed
without calculating $w$ and $c_0$.

We will use a Gaussian $U=e^{-\eta^2}$ as an ansatz function
(\ref{ansatz}). Note that other symmetric bell-shaped functions such
as a {\it{sech}}-function, are possible, too. Then one has $\langle U^n
\rangle = \sqrt{\pi/n}$ and $\langle U_{\eta}^2 \rangle=\langle U^2
\rangle=\sqrt{\pi/2}$. The parameters of the test function $f_0$ and
$w$, and the front velocity $c_0$ can now be determined numerically
using (\ref{u0},\ref{w},\ref{c0}).

Simplifications are possible in the useful limit $\Theta a \ll 1$. In
this limit the (temporal) inverse pulse width $(w_n c_n)^{-1}$ is
small compared to the inhibitor decay time $(\epsilon a)^{-1}$ (see
definition (\ref{theta}). Then in Eqs. (\ref{coeffs}) and (\ref{c0})
the terms proportional to $a$ can be omitted, and for the calculation
of $\langle U^n V \rangle$ one can omit the exponentials in
(\ref{vper},\ref{Vstar}) leading to
\begin{equation}  \label{V-a=0}
V(\eta)=V_s - \int_{0}^{\eta}{U(\eta') d\eta'}\ , 
\quad V_s=\frac{\sqrt\pi}{2} \coth(\frac{a \epsilon L}{2 c_0}) \ .
\end{equation}
Note that $V_s=\sqrt{\pi}/2$ corresponds to $V^\star=0$ (see
(\ref{vper})) for small $a$.  Now $V$ can be replaced by the constant
$V_s$ in $\langle U^n V
\rangle$ (the rest is an odd function) leading to
\begin{equation}  \label{coeffs-a-0}
A=\sqrt\pi(\frac{3}{8} - \frac{5}{6 \sqrt3} \Theta V_s)  \ , \quad 
B=\sqrt\pi(-\frac{5}{6 \sqrt3}(1+u_s)+\frac{1}{\sqrt2} \Theta V_s)  \ , \quad
C=\sqrt{\frac{\pi}{2}} u_s \ ,
\end{equation}
and from (\ref{c0}) we infer
\begin{equation}  \label{epsilon-a-0}
\epsilon=\Theta^2 f_0 \frac{1}{\sqrt6}(1-\frac{1}{\sqrt3} f_0) \ .
\end{equation}

\subsection{Isolated pulses}
We consider now isolated pulses for which the wavelength $L$ is large
compared to the decay length of the inhibitor $1/(\epsilon a)$. In
Figures 1 and 2 we show a comparison of our results
Eqs. (\ref{u0},\ref{w},\ref{c0}) for $f_0,w$ and $c_0$, with a direct
numerical simulation of Eqs. (\ref{barkley}). The pulse shape, the
critical bifurcation point $\epsilon_{c}$ and the behavior near the
saddle-node bifurcation of the amplitude $f_0(\epsilon)$ and of the
velocity $c_0(\epsilon)$ are very well recovered. Note the square-root
behavior near the saddle node.

Let us discuss some systematic features of the isolated pulses at
criticality depending on the equation parameters $a$ and $u_s$, which
can be extracted from our approach. Solutions $f_0$ of (\ref{u0})
exist when the discriminant $B^2-4 A C $ is positive. 
The amplitude at the saddle node $f_c$ is determined by the condition
$B^2-4A C=0$. The corresponding bifurcation parameter $\epsilon_c$ can
then subsequently be determined using Eqs. (\ref{w}) and
(\ref{c0}). Note that the saddle node, which occurs at the maximal
bifurcation parameter $\epsilon_c$, is not given by the relation
$B^2-4 A C =0$
since the condition of maximal $\epsilon$ is different from that of
maximal $\Theta$ (see for example (\ref{epsilon-a-0})). In Fig. 3 we
show $\epsilon_c$ and the corresponding amplitude at the saddle node
$f_c$ as a function of $u_s$ for $a=0$ (continuous line) for the
isolated pulse with $V_s=\sqrt\pi/2$. The points are results from a
full solution of the 1-dimensional version of the ODEs
(\ref{barkleyT1},\ref{barkleyT2}).  In our approximation the limit of
excitability (that is the maximum of $u_s$ for which $\epsilon_c \to
0$) is $u_c=\sqrt{2}(81- 50/\sqrt{2} - 9 \sqrt{81-50
\sqrt{2}})/50=0.4745$, which is close to the exact stall-value $u_c=0.5$. Note
that $u_c$ is independent of $a$, as it should. Note also that pulse
propagation in the neutrally stable case $u_s=0$ is possible. Moreover
pulses are supported even for negative $u_s$ which we have checked
numerically. We mention that for this calculation it is crucial to take
into account the difference between maximizing $\epsilon$ and
maximizing $\Theta$.

In Fig. 4 we show $\epsilon_c$ and $f_c$ as functions of $a$ for
$u_s=0.1$. We see that our approximation (continuous line) reproduces
all features of the ODEs (points). Fig.4 reveals that around $a=1.3$
the saddle-node bifurcation ceases to exist. The velocity $c_0$
approaches zero for these values. This correlates well with the fact
that in the full system pulses become delocalized around $a=1.25$,
i.e. front and tail of a pulse separate creating a domain with $u=1,\
v=1/a$, which represents a locally stable stationary state of the
system. As a matter of fact, the inverse pulse width $w$ diverges here
for the test function approach. \\

\vskip 5pt


\noindent
\subsection {Periodic wave trains}
\vskip 5pt
Even if a given set of equation parameters allows for propagation of a
single solitary pulse, the system may not necessarily support a wave
train consisting of several of such pulses. As a matter of fact, if
the distance $L$ between two consecutive pulses of the train becomes
too small, the pulses run into the refractory tail of the preceding
pulse and consecutively decay. The critical wavelength $L_c$ is a
lower bound for the wavelength for the existence of periodic wave
trains. On can also think of keeping $L$ fixed and, as before, vary
$\epsilon$.  Then the saddle node $\epsilon_c(L)$ is a monotonically
increasing function.

One can calculate $L_c$ (or $\epsilon_c$) essentially as before,
except for the complication that due to $V_s$ (\ref{V-a=0}), the
expressions for $f_0,w$ and $c_0$, (\ref{f_0}),(\ref{w}) and
(\ref{c_0}), cannot be cast in closed form depending only on
$\Theta$. This is true even in the limit of small $a$. However, given
the equation parameters $\epsilon$,$a$ and $u_s$ one may obtain $L_c$
numerically as a consistency relation requiring that at each $L$ there
exists a ${\tilde{\Theta}}=1/(c_0 w)$ so that the value for $c_0$
obtained by solving Eqs. (\ref{u0},\ref{w}) and using the relation
$c_0={\tilde{\Theta}}/w$, matches the value for $c_0$ obtained by
solving Eq. (\ref{c0}). We obtain very good agreement between our test
function approach and the numerically obtained values for the critical
wavelength $L_c$. In Fig. 5 we show a comparison of the values
obtained by integrating the full system (\ref{barkleyT1}) and
(\ref{barkleyT2}) with the calculation of the test function approach
as described above. The critical wavelength $L_c$ diverges when
$\epsilon$ approaches $\epsilon_c$ where the saddle node of the
localized pulse (i.e. $L=\infty$) causes propagation failure of
isolated pulses (see Section 2.1 and Fig. 2).

In the remainder of this section we will discuss the limit of large
values of $L$, i.e. small perturbations to the saddle node
$\epsilon_c$ of the isolated pulse. This causes small shifts of the
critical $\epsilon$, the amplitude $f_0$ and velocity $c_0$ when
compared to their respective values in the case of isolated pulses and
$L=\infty$. For simplicity, we restrict the calculation to the
particular limit of small $a$.

We write
\begin{equation}   \label{Stoer-V_s}
V_s=\frac{\sqrt\pi}{2} \sqrt{1+r}\ , \quad 
r=\sinh^{-2}(\frac{a \epsilon L}{2 c_0}) \approx 4 \exp{(-a \epsilon L/c_0)}
\end{equation}
and expand in terms of small $r$. The correction to the isolated pulse
$r$ is connected to the exponential tail of the inhibitor from the
previous pulse. It thus captures the interaction between pulses
mediated by the inhibitor. At leading order the shift with respect to
the case $L=\infty$ of $\epsilon_c$, $f_0$ and $c_0$ at the saddle
node must be proportional to r. For small $a$ one finds at leading
order
\begin{equation}  \label{Stoer-eps}
(\epsilon_c(\infty)-\epsilon_c(L))/\epsilon_c(\infty)=r=\gamma
\exp{(-a \epsilon L/c_0)} \ , \quad \gamma=4 \ .
\end{equation}

To see this note that the parameters $f_0,w,c_0$ are entirely given as
a function of $R := \Theta V_s$ via Eqs. (\ref{u0}),(\ref{w}) and
(\ref{c0}). Hence, if we multiply Eq. (\ref{epsilon-a-0}) by $V_s^2$,
we have four equations to determine the parameters $f_0,w,c_0$ and
$R$. We expand $f_0,R$ and $\epsilon_c$ with respect to $r$ around the
solutions of the isolated pulse. In particular, we write $\epsilon_c(L) =
\epsilon_c(\infty) + r \epsilon_1$. The first-order correction
$\epsilon_1$ can be entirely determined using
\begin{equation}  \label{epsilon-a-0-r}
\coth^2{\frac{a \epsilon L}{2 c}}
\epsilon=R^2 f \frac{1}{\sqrt6}(1-\frac{1}{\sqrt3} f) \; ,
\end{equation}
where we deliberately ignored the subscripts to denote that $f$ and
$c$ need to be expanded in $r$. The saddle-node condition $d
\epsilon/d f = 0$ implies that the derivative of the right-hand
side of (\ref{epsilon-a-0-r}) with respect to $f$ vanishes at $r=0$,
and we obtain the first-order correction $\epsilon_1$ leading to
(\ref{Stoer-eps}). This result is confirmed by our numerical
simulations. We have determined the shift of $\epsilon_c$ due to
finite wave length $L$ for $a \to 0$ and $u_s=0.1$ numerically from
the ODEs and find $\gamma= 4.3$. This shows the accuracy of our test
function approach for the saddle node shift. The leading-order
approximation for the saddle node behavior is good down to about $a
L=40$. For $a=0.22$ we find numerically $\gamma=5.5$.

In Sec. \ref{summary} we will touch on some questions of stability of
the wave train solutions near the saddle node.



\section{Growing velocity and retracting fingers in two dimensional
excitable media}

In this Section we develop a $2$-dimensional extension of the test
function approach. We study isolated finger solutions, i.e. solutions
which in the $-y$ direction go over into an isolated pulse (moving
with velocity $c_0$ in the $x$ direction) and rapidly decay to zero in
the $+y$ direction.  In the $y$ coordinate they may be regarded as
fronts, which grow or retract with velocity $c_g$ (see Fig. 6 for a
retracting case). We derive an explicit formula for the growing
velocity $c_g$. We now investigate the full $2$-dimensional system
(\ref{barkleyT1}) and (\ref{barkleyT2}) in a frame moving with
velocity $(c_0,\ c_g)$.

We introduce a product ansatz for the activator field
\begin{eqnarray} \label{prod_u}
u(x,y)= f(y)U(\eta)\; .
\end{eqnarray}
with test function $U(\eta)$. This approximation neglects possible
curvature at the tip. Again, we avoid any further uncontrolled
approximation and solve for $f(y)$ and $v(x,y)$ in a systematic
way. Note that $f(y)$ replaces the constant $f_0$ in
(\ref{ansatz}). The solution of Eq. (\ref{barkleyT2}) with the ansatz
(\ref{prod_u}) can be written explicitly as
\begin{equation}  \label{FullSol}
v(\eta,y)=-\Theta \int_{\infty}^{\eta}
  { e^{a \Theta (\eta-s)} f(\frac{1}{\Delta}(\frac{c_g}{c_0 w} (s-\eta)
+y  \Delta )) U(\frac{1}{\Delta}(s+(\frac{c_g}{c_0 w})^2 \eta)) d s} \ 
\end{equation}
where $\Delta=1+(\frac{c_g}{c_0 w})^2$. Further calculations with this
full expression appear prohibitive. Since we are mainly interested in
the reversal point of $c_g$ we resort to an (asymptotic) expansion in
powers of $c_g$, restricting ourselves here to the first two terms.
From (\ref{FullSol}) one obtains
\begin{eqnarray} \label{prod_v}
v(\eta,y)= \Theta \left(g_0(y)V_0(\eta) 
       + c_g g_1(y)V_1(\eta)+{\cal{O}}(c_g^2)\right) \; ,
\end{eqnarray}
where
\begin{eqnarray} \label{g0}
g_0(y)=f(y) \qquad {\rm{and}} \qquad g_1(y)=f^\prime(y)\; .
\end{eqnarray}
$V_0(\eta)$ coincides with $V(\eta)$ of Eq. (\ref{vper}) with
$V^\star=0$ and $L=\infty$ (or, for small $a$, with Eq. (\ref{V-a=0}))
and
\begin{eqnarray} \label{V1}
V_1(\eta)=-\frac{1}{c_0w}e^{a\Theta
\eta}\int_{\infty}^{\eta}e^{-a\Theta \eta^\prime}V_0(\eta^\prime)\,
d\eta^\prime\; .
\end{eqnarray}
Note that the first-order correction of the inhibitor (\ref{V1}) can
also be obtained by inserting the ansatz (\ref{prod_v}) into
(\ref{barkleyT2}) and solving for successive orders of $c_g$.

Repeating the procedure that led to Eq. (\ref{projectionsU}) for
$f_0$, i.e. multiplying Eq. (\ref{barkleyT1}) with $U(\eta)$ and
integrating over $\eta$, we obtain using (\ref{g0})
\begin{eqnarray}    \label{cubic}
D f'' + c_g f' + F(y) f=0
\end{eqnarray}
with
\begin{eqnarray}
F(y)=[-D w^2 \langle U_{\eta}^2\rangle +\langle U^2(1-f (y)U)
[f(y)U-\Theta(f(y)V_0+c_gf^\prime(y)V_1)-u_s]\rangle /\langle U^2\rangle . 
\end{eqnarray}
Note that for $f(y) \equiv f_{0\pm}$, where $f_{0\pm}$ are the
solutions of the quadratic equation (\ref{u0}), we have $F(y)=0$.

We first neglect the higher-order correction of the inhibitor field
$V_1(y)$. Then $F$ is a quadratic form in $f(y)$ with zeros
$f_{0\pm}$, and Eq. (\ref{cubic}) can be solved exactly with the
ansatz
\begin{eqnarray} \label{sneyd}
f^\prime(y) =\alpha  f(f-f_0)\;,
\end{eqnarray}
which states that far away from the tip $f(y)$ is constant and takes
values $0$ or $f_0$. The constant of proportionality $\alpha$ can be
determined. Using (\ref{sneyd}) we obtain for the growing velocity
\begin{eqnarray}
\label{cg0}
{c_{g0}}=\sqrt{\frac{D}{2}}
      \sqrt{\frac{\langle U^4\rangle - \Theta\langle U^3V \rangle}
                 {\langle U^2 \rangle}} (f_{0+}-2f_{0-})\;.
\end{eqnarray}
The point $c_{g0}=0$ is fixed by the condition $f_{0+}=2 f_{0-}$. The
value for $\epsilon$ where $c_{g0}=0$ which we denote by $\epsilon_g$,
matches very well the value obtained by numerically integrating the
full $2$-dimensional system (\ref{barkleyT1}) and
(\ref{barkleyT2}). However, the behavior for nonzero growing
velocities is not captured by (\ref{cg0}). In fact, the slope
$\partial {c_{g0}}/\partial \epsilon$ close to the reversal point is
too small by an order of magnitude for the parameters used in
Fig. 1,2,5,6 when compared to the values obtained by the full
$2$-dimensional simulation.

To obtain the correct slope we need to take into account the
correction $V_1(\eta)$ of the inhibitor field.  Including the
first-order correction $V_1(\eta)$ in Eq. (\ref{cubic}) we solve for
$c_g$ analogously to the determination of $c_0$ in Section 2 by
multiplying Eq. (\ref{cubic}) by $f^\prime(y)$ and subsequently
integrating over $y$. To ${\cal{O}}(c_g)$ we obtain
\begin{eqnarray}
\label{CG}
c_g=c_{g0} \frac{1}{1+\frac{1}{2}G_0f_o+\frac{3}{10}G_1f_o^2}\; ,
\end{eqnarray}
where 
\begin{eqnarray}
\label{G0G1}
G_0=-\Theta\frac{\langle U^2V_1\rangle}{\langle U^2 \rangle} 
\qquad  {\rm{ and}} \qquad 
G_1=\Theta\frac{\langle U^3V_1\rangle}{\langle U^2 \rangle} \; .
\end{eqnarray}
As expected the higher-order corrections $G_0$ and $G_1$ do not change
the value of $\epsilon_g$, but change the slope of $\partial c_g
/\partial \epsilon$. In Fig. 7 we show a comparison of the test
function approach and of Eq. (\ref{CG}) with numerically obtained
data. The correspondence close to $c_g=0$ is striking. To obtain
better agreement further away from $c_g=0$ one would have to include
higher-order terms in the expression (\ref{prod_v}).\\




\section{Summary and discussion}  \label{summary}
We have developed a non-perturbative method to study critical wave
propagation of single pulses and periodic wavetrains in $1$ and $2$
dimensions. The method is based on the observation that near the
bifurcation point the pulse shape is close to a symmetric bell-shaped
function. A test function approximation, optimizing the two free
parameters of a bell-shaped function, i.e. amplitude and width, allows
to calculate the wave speed of a critical and close-to-critical
pulse. We were able to study propagation failure of isolated pulses
and wave trains, and moreover the test is capable of capturing more
general features such as the transition from excitability to
bistability. We have also performed our test function method with a
more general class of nonsymmetric test functions to calculate the
pulse parameters and velocities. It turns out that near the saddle
node the asymmetry indeed becomes irrelevant. 

We extended our method to two dimensional situations, and used it to
study broken fronts. Depending on parameters these fingers may either
retract or sprout and start spiraling. We studied the growing velocity
of a critical finger whose growing velocity is close to zero. Our test
function method combined with a separation ansatz for the
two-dimensional finger tip yields analytical expressions for the
growing velocity which depend only on the equation parameters and
which are for large parameter ranges in good agreement with the
numerically obtained values. We elaborated on the importance of the
inhibitor field for the growing velocity.

Let us finally mention an interesting observation for one-dimensional
wave trains. Below some (rather large) pulse separation (wavelength)
the wave train looses stability not via a saddle node but via a Hopf
bifurcation. This may be seen in numerical simulations of the full
system (\ref{barkleyT1}) and (\ref{barkleyT2}) (and also of other
excitable media equations such as the Fitzhugh-Nagumo equation) and in
analytical approximations, either based on a time-dependent
generalization of the test function approach or on a systematic
reduction scheme valid near the saddle node. At the codimension $2$
point where the saddle node of the wave train coincides with the
Hopf-bifurcation the Hopf frequency becomes infinite. This
time-dependent phenomenon can of course not be captured within the
current stationary algebraic framework. This phenomenon and its
implications will be published separately. For completeness we have
included a time-dependent extension of the test function approach
which allows for Hopf bifurcation in an Appendix. this seems to be
related to the phenomenon of alternans \cite{Nolasco} which has been
studied in a different parameter regime far away from the saddle node
\cite{Karma_A}.

Another interesting problem we plan to address is the selection of the
wavelength (or pitch) of spirals. In simulations we found that near
$\epsilon_g$, where growth of fingers becomes small, the selected
wavelength diverges. Kinematic theory
\cite{BCF,Zykov,Mikhailov1,Tyson} addresses this problem. Kinematic
theory provides, in principle, a relationship between the rotation
frequency of a spiral and its core radius. However, it fails to
provide expressions for either of the two which only depend on the
equation parameters. A connection of our theory with kinematic theory
is planned for further research to fill this gap.\\


{\underbar{\bf Acknowledgements }} We would like to thank the INLN
where parts of this work was done for its hospitality. G.G. would like
to thank Valentin Krinsky for introducing him to this research. We
would like to thank Dwight Barkley, Gianne Derks, Alain Karma, Jim
Keener, Ian Melbourne, Sasha Panfilov, Alain Pumir, Sebastian Reich
and Mark Roberts for valuable discussions. G.G. was partly supported
by a European Commission Grant, contract number HPRN-CT-2000-00113,
for the Research Training Network {\it Mechanics and Symmetry in
Europe\/} (MASIE) and by an Australian Academy of Science Travel
Grant.  Part of the simulations were done with the XDim Interactive
Simulation Package developed by P. Coullet and M. Monticelli, who
L. K.  thanks for their support.

\vskip 5pt
\vfill\eject


\appendix 
\section{Appendix: Time-dependent system for wave propagation of
pulses and periodic wave trains in one-dimensional excitable media} In
this appendix we present the time-dependent calculation for our test
function approach. For simplicity we restrict ourselves to small
values of $a$. This means that the width of the activator pulse $u$ is
small compared to the width of the decaying inhibitor field $v$. We
include now temporal dependency of the pulse variables $f_0,c_0$ and
$w$. We study pulse trains and note that the localized pulse can be
obtained in the limit $L\to
\infty$. We hence choose $u$ of the form
\begin{eqnarray}  
\label{ansatzT}
u(x)=\sum_0^N f_n(t) U(\eta_n)\; .
\end{eqnarray}
The sum extends over all $N$ pulses. We defined
$\eta_n=w_n(t)(x-\phi_n(t))$ and $U(\eta)$ is chosen as a Gaussian
$e^{-\eta^2}$ as above. We allow for individual dynamics of the pulses
characterized by the amplitude $f_n(t)$, the inverse width $w_n(t)$,
and the position $\phi_n(t)$. For a stationary wavetrain the $f_n$ and
$w_n$ will be constant and all equal, and $\phi_n= n (c/L) t$, where
$c$ is the velocity. We will restrict ourselves to the situation where
the perturbations around such a state are small and slowly varying.
We insert the ansatz (\ref{ansatzT}) into the general expression
$$v(t,x)=-\epsilon
\int_{\infty}^t e^{-\epsilon a(t'-t)} u(t',x)\;, d t^\prime$$ obtained from the
second Eq. (\ref{barkley}). Assuming that the (temporal) inverse pulse
width $(w_n c_n)^{-1}$ is small compared to the inhibitor decay time
$(\epsilon a)^{-1}$, we obtain the following expression valid in the
vicinity of the $n=0$ pulse, which is assumed to pass the origin at
$t=0$
\begin{eqnarray} 
\label{vT} 
&&v(\eta_0)= \epsilon
\frac{f_0}{w_0 c_0}[\frac{1}{2} \langle U\rangle -\int_0^\eta d \eta'
{U(\eta')} ] 
+\epsilon \langle U\rangle
\sum_{l=1}^{\infty} e^{-l \epsilon a L/c_l} \frac{f_l}{w_l c_l} +s_0
\int_0^\eta d \eta' {\eta' U(\eta')} 
\end{eqnarray} 
where the brackets indicate integration over the whole $\eta$-domain,
and
\begin{eqnarray} 
\label{s_0} 
s_0=-\epsilon (\partial_t[f_0/(c_0 w_0)])/(w_0 f_0).  
\end{eqnarray}
        
In order to determine $f_n,\ w_n$, and $c_n$ we again project Eq.
(\ref{barkleyT1}) onto $U(\eta)$, $\eta U'(\eta)$, and
$U'(\eta)$. After combining the first two equations appropriately one
obtains
\begin{eqnarray}    \label{f_0}
&&\langle U^2\rangle (\frac{\dot f_0}{f_0}-\frac{\dot w_0}{w_0})=-u_s \langle U^2\rangle 
+\frac{5}{6} (1+u_s) \langle U^3\rangle  f_0 
 -\frac{3}{4} \langle U^4\rangle  f_0^2  \nonumber \\
&&- (\langle U^2\rangle -\frac{5}{6} \langle U^3\rangle f_0) V_p
-(\frac{5}{6} \langle U^3\rangle -\frac{3}{4}\langle U^4\rangle f_0) \frac{s_0}{2}  \ , \\ 
         \label{w_0}
&&\langle U^2\rangle  \frac{\dot w_0}{w_0}= 2 D w_0^2\langle U'^2\rangle +\frac{1}{3}
(1+u_s) \langle U^3\rangle  f_0-\frac{1}{2}\langle U^4\rangle f_0^2  \nonumber \\
&&+\frac{1}{3}\langle U^3\rangle  f_0 V_p
-(\frac{1}{3} \langle U^3\rangle -\frac{1}{2}\langle U^4\rangle f_0) \frac{s_0}{2} \ ,
\end{eqnarray}
where
\begin{eqnarray}   \label{V_p}
        V_p=\epsilon \langle U_1\rangle (\frac{1}{2} \frac{f_0}{c_0 w_0}+ 
\sum_{l=1}^{\infty} e^{-l \epsilon a L/c_l} \frac{f_l}{w_l c_l})\  .
\end{eqnarray}
The third projection gives an algebraic relation
\begin{eqnarray}   \label{c_0}
c_0 w_0\langle U'^2 \rangle=\frac{\epsilon f_0}{c_0 w_0}
(\frac{1}{2}\langle U^3\rangle  - \frac{f_0}{3}\langle U^4\rangle )  \; ,
\end{eqnarray}
These equations are written for the pulse $n=0$. They apply correspondingly
to the other pulses.

First consider the case of a stationary, isolated pulse ($L=\infty$)
where $s_0=0$ and the sum in Eq. (\ref{V_p}) vanishes.  The resulting
algebraic relations are easily solved, and correspond in the limit of
$a\to 0$ to Eqs. (\ref{u0}), (\ref{w}), (\ref{c0}).  With reasonable
initial conditions, simulation of the ODEs (\ref{f_0}), (\ref{w_0}),
together with (\ref{c_0}) leads to the same result as described in
Section 2. The ODEs (\ref{f_0},\ref{w_0}) relax to the stationary
values obtained in Section 2.\\

This time-dependent test function approach allows to go beyond the
stationary bifurcations discussed in Section \ref{summary}. As
mentioned in Section 2 the saddle-node bifurcation related to
propagation failure for well separated pulses transforms into a
subcritical Hopf bifurcation via a Takens-Bogdanov point when the
pulse separation is reduced below a critical value $L_c$. This work
will be published elsewhere. The Hopf-bifurcation can be captured
within the ODE-system (\ref{f_0}), (\ref{w_0}) and (\ref{c_0}), and
some preliminary simulations have been done.


\newpage

 \begin{figure}[htb]
   \begin{center}
    \includegraphics[angle=0,width=.8\textwidth]{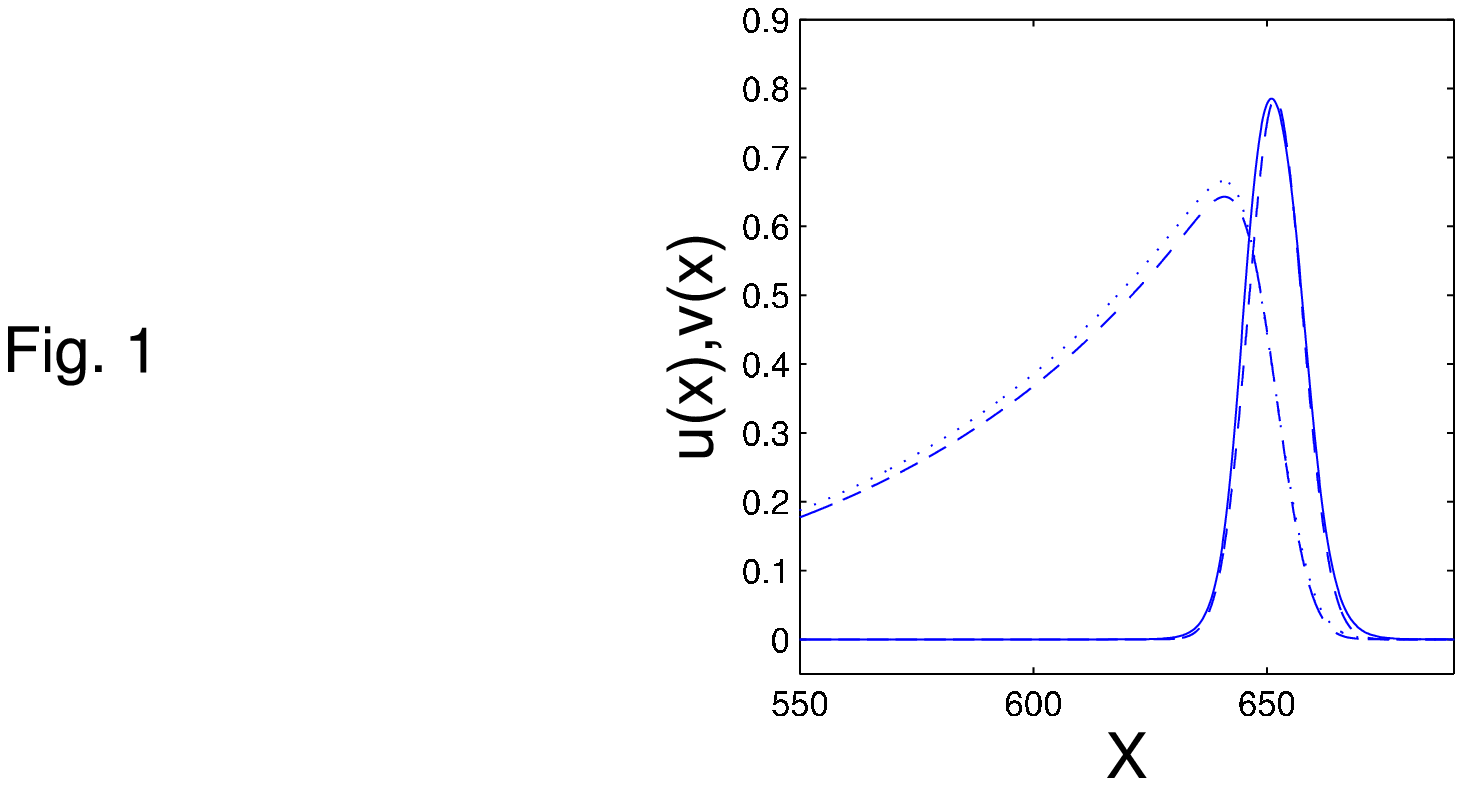}
    \end{center} 
   \caption{ Activator field $u$ of a steady front with
   $\epsilon\approx\epsilon_c$ at $\epsilon=0.0485$. The points depict
   the solution obtained by numerically integrating
   Eqs. (\ref{barkley}). The continuous line is the theoretical curve
   obtained with the test function approach. Note that here $\epsilon$
   is not even very close to $\epsilon_c \approx 0.049$.}
 \end{figure}

 \begin{figure}[htb]
   \begin{center}
    \includegraphics[angle=0,width=.8\textwidth]{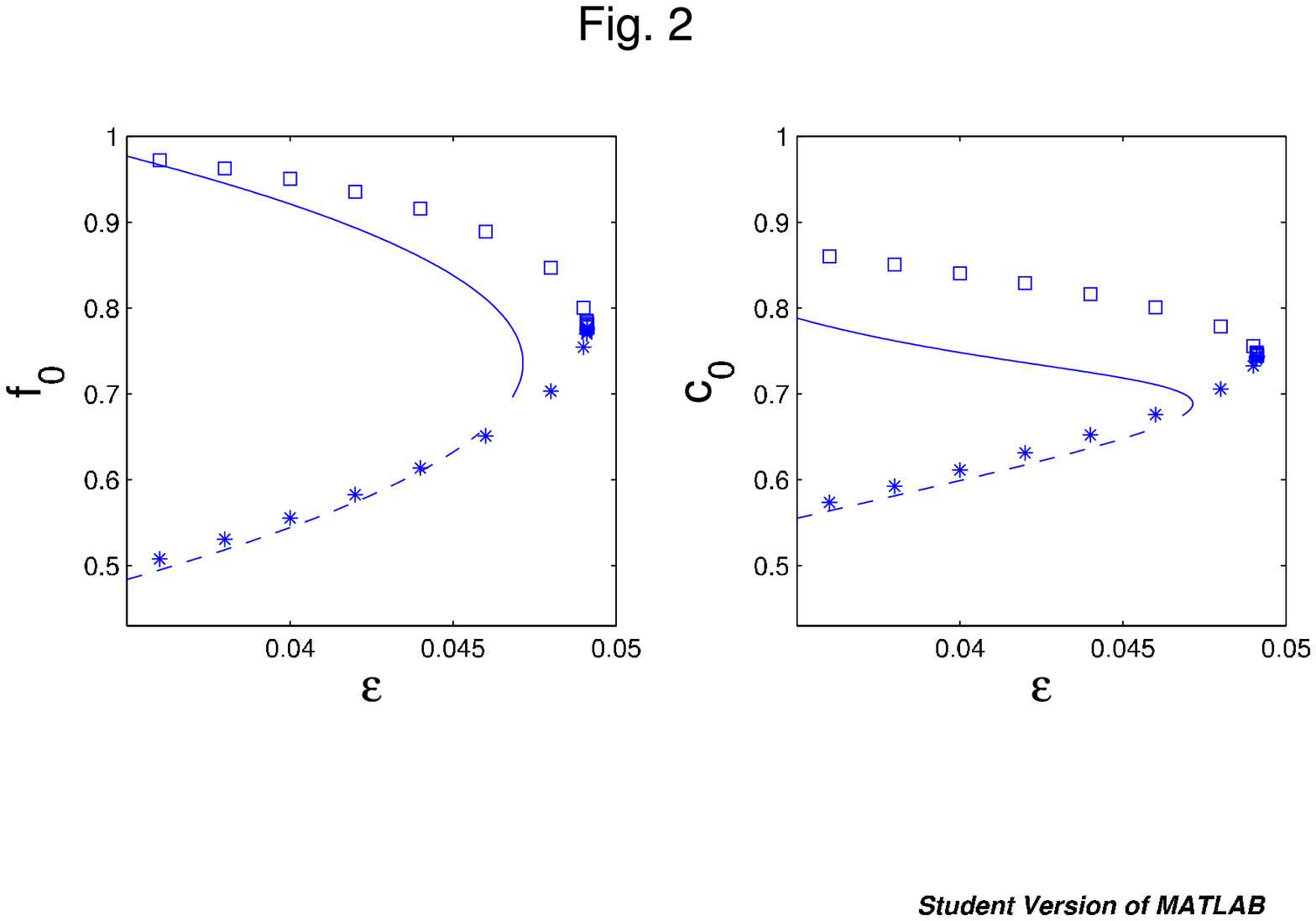}
    \end{center} 
   \caption{ (a): Amplitude $f_o$ as a function of
$\epsilon$. (b): Velocity $c_0$ as a function of $\epsilon$. The
parameters used were $D=3.0$, $a=0.22$ and $u_s=0.1$. The crosses
depict the numerically obtained values of integrating the full system
(\ref{barkleyT1}) and (\ref{barkleyT2}), the lines depict the stable
and unstable branch calculated with the test function approach,
i.e. by using Eqs. (\ref{u0},\ref{w},\ref{c0}).}
 \end{figure}

 \begin{figure}[htb]
   \begin{center}
    \includegraphics[angle=0,width=.8\textwidth]{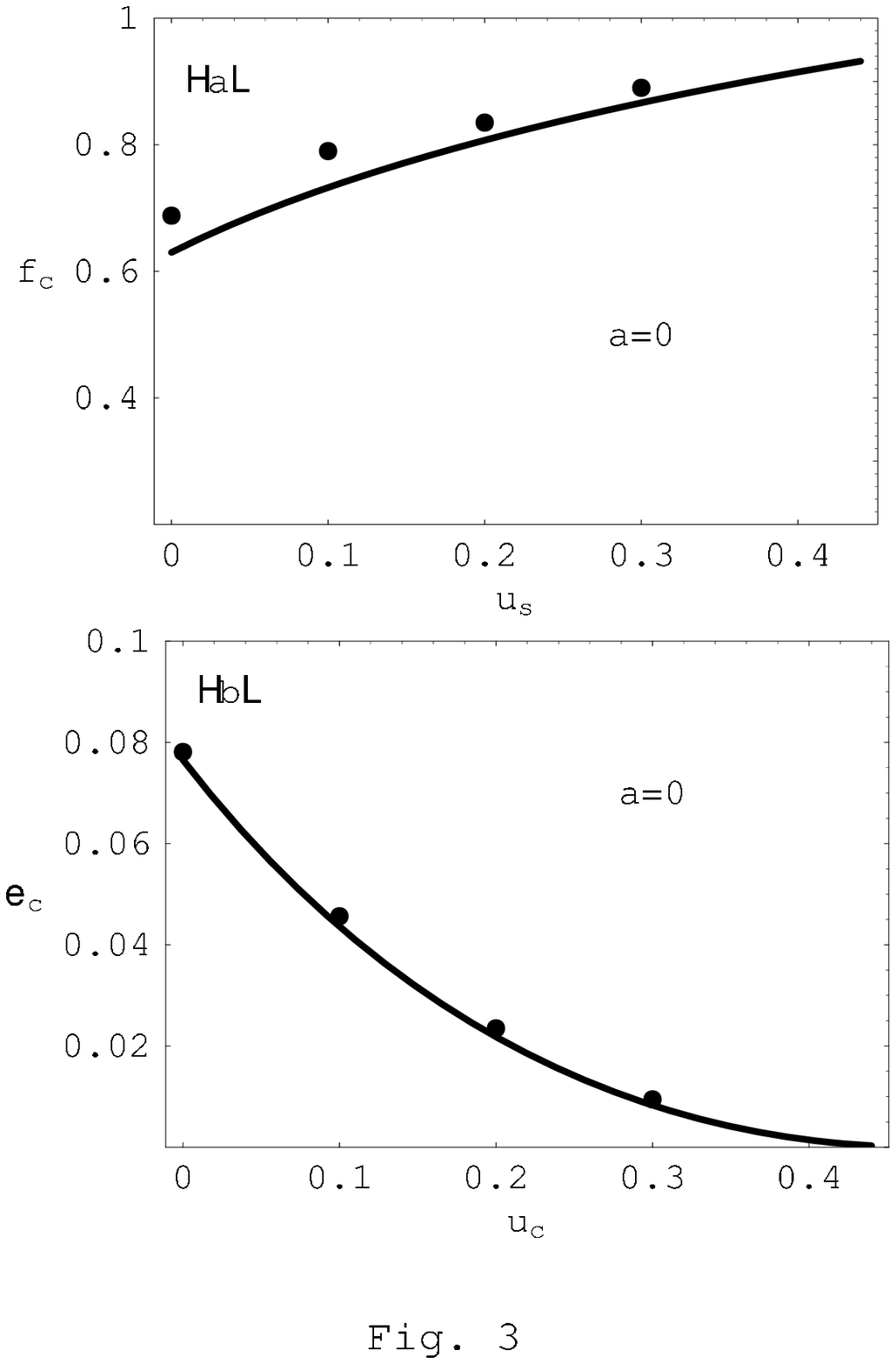}
    \end{center} 
   \caption{Behavior at the saddle node of an isolated pulse
for $a=0$. (a): The critical $f_c$ versus $u_s$. (b): Critical
amplitude $\epsilon_c$ versus $u_s$. Points correspond to simulations of the
full $1$-dimensional version of (\ref{barkleyT1},\ref{barkleyT2}). The
continuous line corresponds to the test function approach.}
 \end{figure}

 \begin{figure}[htb]
   \begin{center}
    \includegraphics[angle=0,width=.8\textwidth]{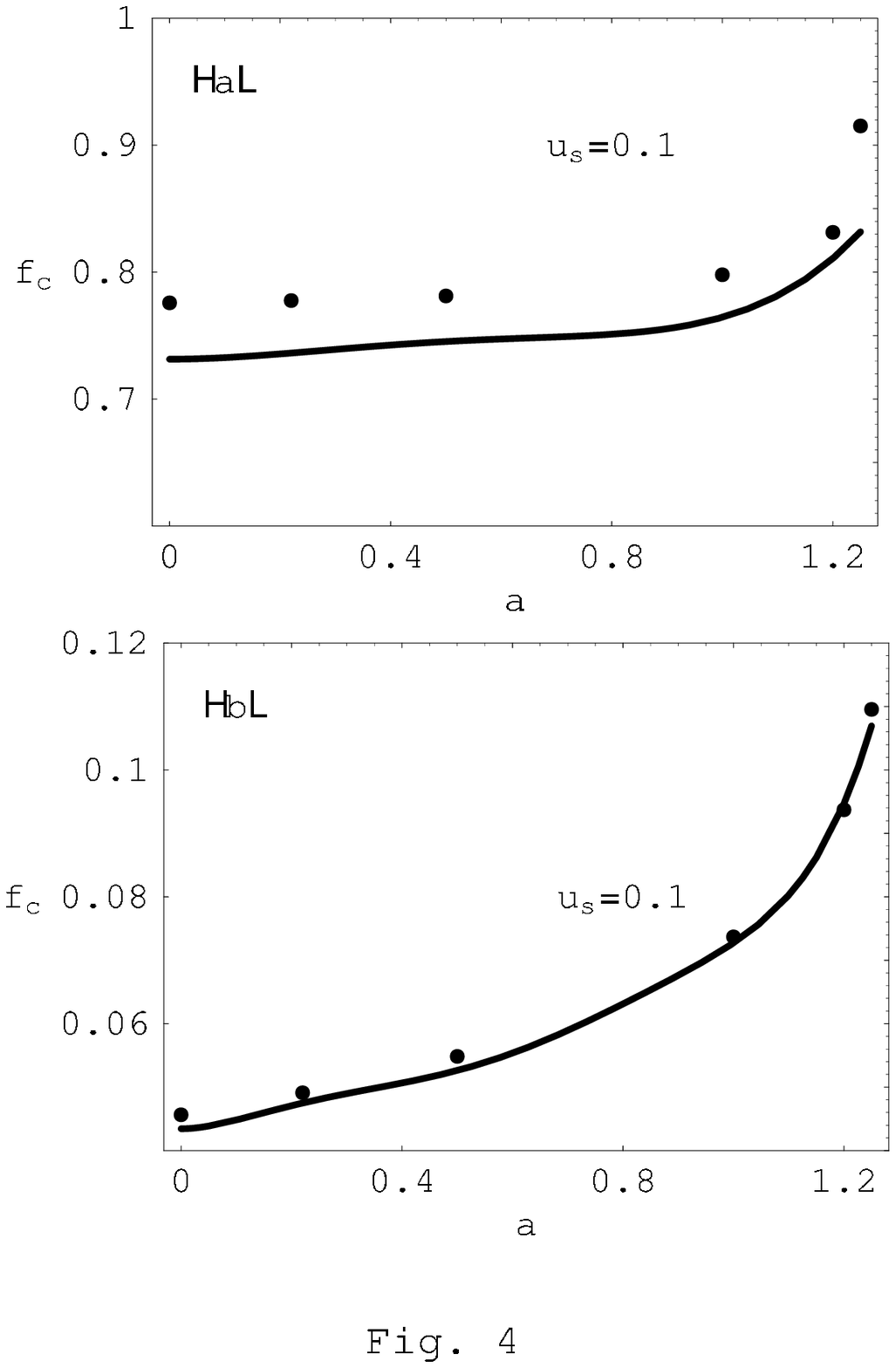}
    \end{center} 
   \caption{Behavior at the saddle node of an isolated pulse
for $u_s=0.1$. (a): The critical $\epsilon_c$ versus $a$. (b): Critical
amplitude $f_c$ versus $a$. Points correspond to simulations of the
full $1$-dimensional version of (\ref{barkleyT1},\ref{barkleyT2}). The
continuous line corresponds to the test function approach.}
 \end{figure}

 \begin{figure}[htb]
   \begin{center}
    \includegraphics[angle=0,width=.8\textwidth]{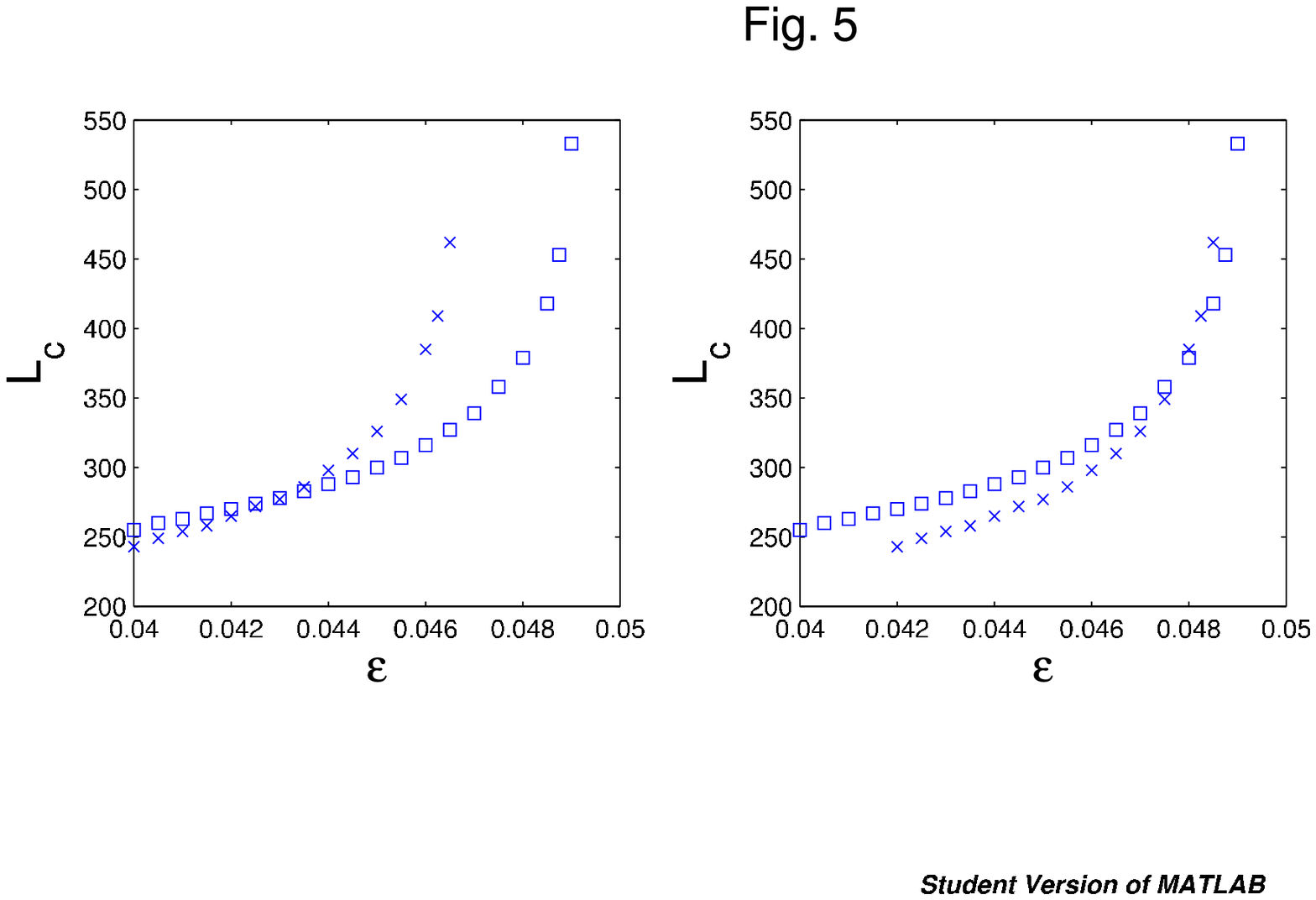}
    \end{center} 
   \caption{(a) The critical wavelength $L_c$ as a function
of $\epsilon$. Crosses depict the values obtained by direct simulation
of the full system (\ref{barkleyT1}) and (\ref{barkleyT2}). Squares
depict the values obtained by the test function approach described in
Section 3. Parameters are again $D=3.0$, $a=0.22$ and $u_s=0.1$. In
(b) the same numerical results as in (a) are presented but here the
data-points corresponding to the test-function approach are shifted
along the $\epsilon$-axis such that the saddle nodes at
$L_c=\infty$ (see Fig. 2), coincide.}
 \end{figure}

 \begin{figure}[htb]
   \begin{center}
    \includegraphics[angle=0,width=.8\textwidth]{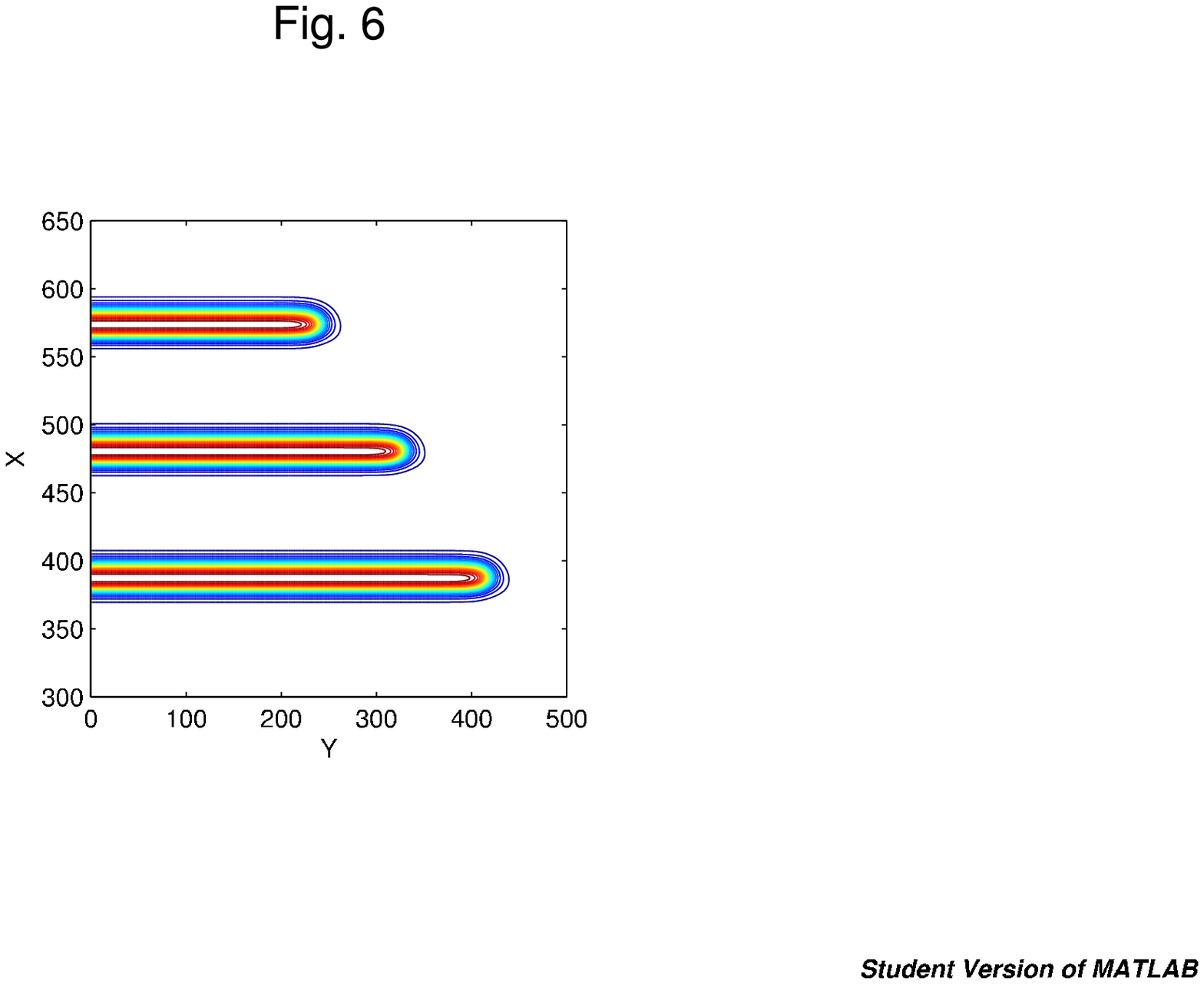}
    \end{center} 
   \caption{Contour plot of the activator $u$ of a retracting
finger at different times.}
 \end{figure}

 \begin{figure}[htb]
   \begin{center}
    \includegraphics[angle=0,width=.8\textwidth]{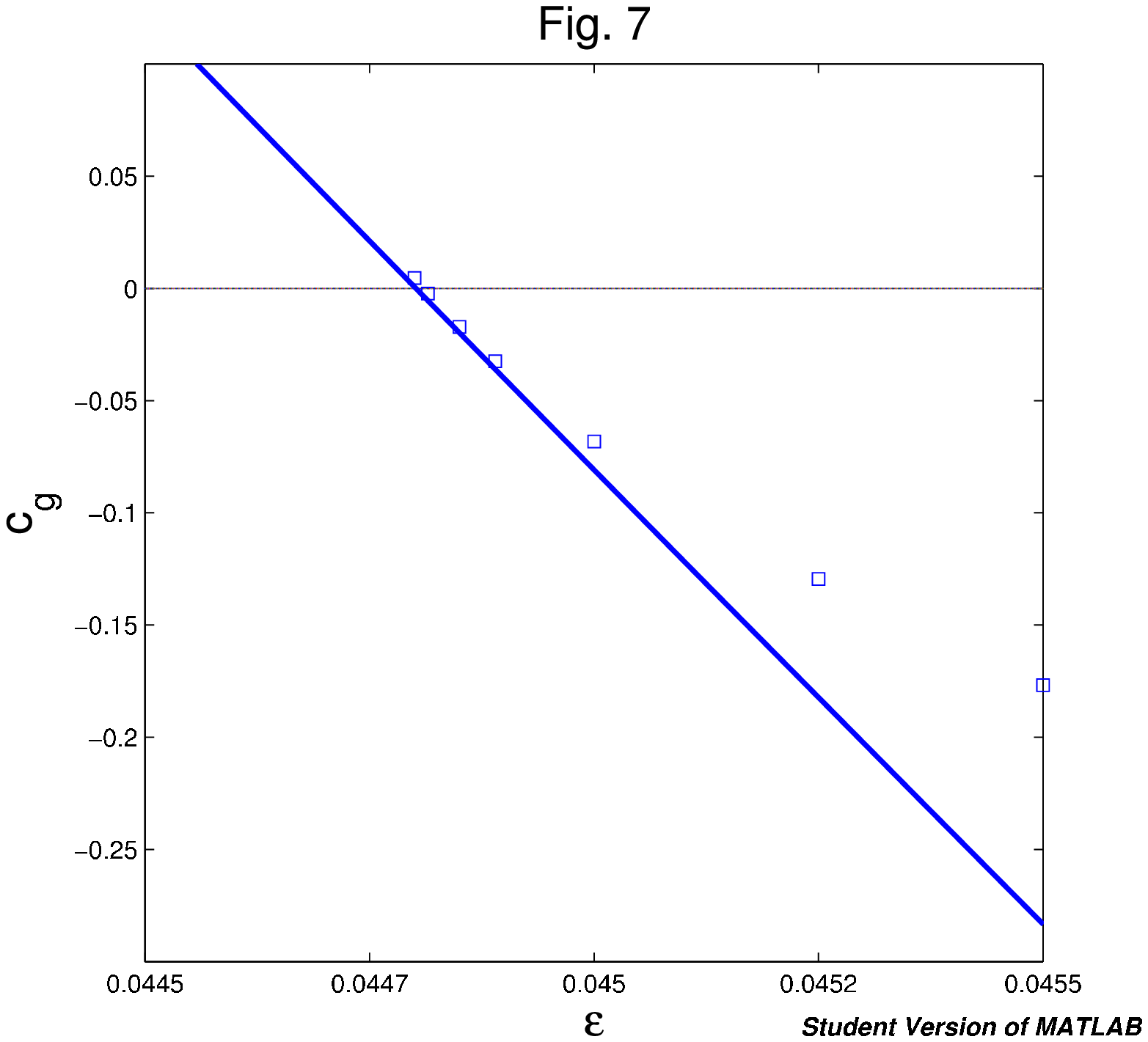}
    \end{center} 
   \caption{Growing velocity $c_g$ as a function of
$\epsilon$. The crosses depict the values obtained by direct numerical
integration of Eqs. (\ref{barkleyT1}) and (\ref{barkleyT2}). The
continuous line shows the theoretical curve (\ref{CG}) using the test
function approach of Section 4.1. We have shifted the curve along
the $\epsilon$-axis by the difference of the $\epsilon$-values $\Delta
\; \epsilon=0.001972$ for the saddle nodes obtained by the numerical
simulations of the full system (\ref{barkleyT1}) and
(\ref{barkleyT2}), and the test function approach of Section 2.
}
\end{figure}

\end{document}